\documentclass[%
aip,
amsmath,amssymb,
reprint
]{revtex4-1}

\usepackage{graphicx}% Include figure files
\usepackage{dcolumn}% Align table columns on decimal point
\usepackage{bm}% bold math
\usepackage[section]{placeins}

\usepackage{xpatch}
\usepackage{xcolor}
\makeatletter
\ExplSyntaxOn
\cs_new:Npn \bibColoredItems #1#2
  {
    \clist_map_inline:nn {#2} { \cs_new:cpn {bib@colored@##1} {#1} }
  }
\ExplSyntaxOff

\newcommand\bib@setcolor[1]{%
  \ifcsname bib@colored@#1\endcsname
    \expandafter\color\expandafter{\csname bib@colored@#1\endcsname}
  \else
    \normalcolor
  \fi
}

\xpatchcmd\@bibitem
  {\item}
  {\bib@setcolor{#1}\item}
  {}{\fail}

\xpatchcmd\@lbibitem
  {\item}
  {\bib@setcolor{#2}\item}
  {}{\fail}
\makeatother

\usepackage{endnotes}

\usepackage{amsmath}
\usepackage{color}
\usepackage[normalem]{ulem}
\usepackage{balance}
\usepackage[T1]{fontenc}

\newcommand{\mycite}[1]{\scalebox{1.3}[1.3]{\raisebox{-0.80ex}{\cite{#1}}}}
\begin{document}

\preprint{AIP/123-QED}

\title{Random-walk shielding-potential viscosity model for warm dense metals}% Force line breaks with \\

\author{Yuqing Cheng}
\affiliation{School of Mathematics and Physics, University of Science and Technology Beijing, Beijing 100083, China}
\affiliation{Laboratory of Computational Physics, Institute of Applied Physics and Computational Mathematics, Beijing 100094, China}%

\author{Haifeng Liu}\thanks{Corresponding author: liu\_haifeng@iapcm.ac.cn}
\affiliation{Laboratory of Computational Physics, Institute of Applied Physics and Computational Mathematics, Beijing 100094, China}%
\author{Yong Hou}
\affiliation{Department of Physics, College of Liberal Arts and Sciences, National University of Defense Technology, Changsha 410073, China}%
\author{Xujun Meng}
\author{Qiong Li}
\author{Yu Liu}
\author{Xingyu Gao}
\affiliation{Laboratory of Computational Physics, Institute of Applied Physics and Computational Mathematics, Beijing 100094, China}%
\author{Jianmin Yuan}
\affiliation{Graduate School, China Academy of Engineering Physics, Beijing 100193, China}%
\author{Haifeng Song}
\author{Jianguo Wang}
\affiliation{Laboratory of Computational Physics, Institute of Applied Physics and Computational Mathematics, Beijing 100094, China}%

%\date{\today}% It is always \today, today,
             %  but any date may be explicitly specified

\begin{abstract}
We develop a novel model, called the ``random-walk shielding-potential viscosity model'' (RWSP-VM) that introduces the statistics of random-walk ions and the Debye shielding effect to describe the viscosities of warm dense metals. The viscosities of several metals with low to high atomic number (Be, Al, Fe, and U) are calculated using the analytical expression of RWSP-VM. Additionally, we simulate the viscosities of Fe and Be by employing the Langevin molecular dynamics (MD) and classical MD, while the MD data for Al and U are obtained from a previous work. The results of the RWSP-VM are in good agreement with the MD results, which validates the proposed model. Furthermore, we compare the RWSP-VM with the one-component plasma model and Yukawa viscosity model and show that the three models yield results in excellent agreement with each other in the regime where the RWSP-VM is applicable. These results indicate that the RWSP-VM is a universal, accurate, and highly efficient model for calculating the viscosity of metals in the warm dense state. The code of the proposed RWSP-VM is provided, and it is envisaged that it will have broad application prospects in numerous fields.
\end{abstract}

%\keywords{Suggested keywords}%Use showkeys class option if keyword
                              %display desired
\maketitle

%\tableofcontents

\section{\label{sec:Introduction}Introduction}
%\textbf{\textit{Introduction}}.---
The transport properties of matter have been widely investigated. In particular, the shear viscosity over a wide parameter space is crucial for designing inertial confinement fusion (ICF) targets \cite{appLindl}, understanding wave damping in dense plasmas \cite{appWong}, microjetting during shock loading \cite{appDurand}, determining the Rayleigh--Taylor instability \cite{appRT1,appRT2}, and understanding the evolution of astrophysical objects \cite{appDai}.
Viscosity measurements have been conducted at pressures close to 1 bar; however, it is very difficult to conduct such measurements at both high temperature and high pressure \cite{extraMiller,extraMa}. Researchers have been investigating the viscosity of materials through three different types of methods.

Firstly, Alf{\` e} and Gillan \cite{methodsAlfe} employed first-principles molecular dynamics (FPMD) simulations to calculate the viscosity of liquid Al and Fe--S using the Green--Kubo relations \cite{methodsAllen}, and several researchers studied the viscosity of liquid Al and Pu using a similar method \cite{methodsWang,extraWang}. In these works, the time step and simulation length were several fs and from dozens of ps up to 100 ps, respectively. However, the number of Kohn--Sham orbitals that ensures the accuracy of the simulation increases rapidly with increasing temperature \cite{extraBlanchet} for warm dense matter, which contains partially ionized electrons that interact strongly with the nuclei. Obtaining a sufficient amount of data remains challenging due to the expensive overhead of the FPMD simulation for warm dense matter.

Secondly, several researchers have calculated the viscosities of hot dense Al and U  using both the Langevin MD (LMD) and classical MD (CMD) simulations based on the electronic and ionic structures derived from  the average-atom model combined with the hyper-netted chain (AAHNC) approximation \cite{methodsHou, dataU}. The results agree well with those of the orbital-free MD (OFMD) simulations and those of the effective potential theory + average atom (EPT+AA) and pseudo-atom MD (PAMD) simulations reported in Refs. \mycite{extraKress}, \mycite{methodsStarrett} and \mycite{methodsDaligault}. In our opinion, the FPMD, LMD, and CMD simulations are the more fundamental calculations because they have the most physics fidelity. Nevertheless, the efficiency of these simulation methods limits their applications.

Thirdly, several researchers have developed practical models to calculate the viscosity \cite{modelsDaligault, modelsMurillo}.
Murillo proposed a model named the Yukawa viscosity model (YVM) to compute the viscosity of materials from the liquid to the warm dense state by mapping the real matter onto the Yukawa model, for which the viscosity can be written in a quasiuniversal form \cite{modelsMurillo}. He employed several dimensionless quantities, such as the coupling parameter $\Gamma$ and screening parameter $\kappa$, to build the model. The data obtained from MD simulations \cite{modelsSaigo}, namely $0.1 \le \kappa \le 3$ and $\Gamma \ge 2$, were used to fit the YVM coefficients.
Daligault et al. developed the one-component plasma (OCP) model to evaluate the viscosity of one-component materials from the weakly-coupled regime to the solidification threshold \cite{modelsDaligault}. They proposed a practical expression for $\kappa=0$, the coefficients of which were derived from the fitting of their equilibrium MD simulations. The applicability regime of these models is usually restricted to the parameter space of the MD simulations.

In this study, we develop a novel model called the ``random-walk shielding-potential viscosity model'' (RWSP-VM) to determine the viscosity of metal elements in the warm dense state without any MD simulations. The viscosities of Be, Al, Fe, and U calculated by the RWSP-VM are consistent with those of Fe and Be obtained from the LMD and CMD simulations in this work and those of Al and U derived in previous works \cite{methodsHou,dataU}. By quantitatively comparing the results of several different models, including the RWSP-VM, OCP, and YVM, it is shown that these three models yield viscosity values in excellent agreement with each other for metal elements in the warm dense state.

\section{\label{sec:Model}Model}
%\textbf{\textit{Model}}.---
For metals in the warm dense state, electrons are partially ionized, and  ion velocities are very high. This indicates that these ions move randomly as an ``ion gas'' driven by the Coulomb interactions between each other. Furthermore, the Debye shielding effect caused by the numerous ionized electrons weakens these Coulomb interactions. The random motion of ions can be treated by employing the random-walk concept \cite{randomwalk}.

Several researchers have utilized this concept to build viscosity models. Stanton and Murillo developed the Stanton--Murillo transport (SMT) model, which is based on the Boltzmann equation (BE), to calculate all transport coefficients, including the viscosity \cite{transportMurillo}. The SMT regimes of the viscosity are relevant to high-energy-density matter. They evaluated the momentum-transfer cross section in the BE using the Coulomb logarithm, which is defined by the screening length. However, it is not easy to accurately estimate the screening length of a material.

In our model, we take into account both the random walk of the ion gas and the Debye shielding effect and make two assumptions.

Assumption 1: The ions move randomly as an ion gas, and only binary collisions of ions are considered for simplicity.

Assumption 2: There is a cutoff distance $r_0$, i.e., only when the distance $r$ between a pair of ions is less than $r_0$, the Coulomb interaction is sufficiently strong to give rise to the shielding potential.

Here, the Coulomb interaction is described as $\phi(r)=q^2/r$, where $q^2=(Ze)^2/(4\pi \varepsilon_0)$, $Ze$ is the charge of the ion, $e$ is the elementary charge, and $\varepsilon_0$ is the vacuum permittivity.
The reason why we employ the Coulomb interaction combined with the cutoff distance rather than an actual Debye-screened potential is that our treatment of the Debye-screened potential is easy to analyze from a mathematical perspective, resulting in a more simplified expression and more accurate results; furthermore, this is a reasonable approximation.

At the beginning, the shear viscosity $\eta$ is described as \cite{methodsAllen}:
\begin{equation}
\eta=\frac{1}{2t}\frac{V}{k_{B}T} {\langle ( \mathcal{L}_{\alpha\beta}(t)-\mathcal{L}_{\alpha\beta}(0) )^2 \rangle},
\label{eq:definition}
\end{equation}
where $t$ is the time, $V$ is the volume of the system, $k_B$ is the Boltzmann constant, and $T$ is the temperature, $\langle \cdots \rangle$ stands for the ensemble average, which varies with time $t$, and $\mathcal{L}_{\alpha\beta}=\frac{1}{V} \sum\limits_i r_{i\alpha}p_{i\beta}$.
It is worthy noticing that the time $t$ should be sufficiently large to ensure a physically reasonable ensemble average.
Here, $\alpha$ and $\beta$ are the directions in the Cartesian coordinate system with $\alpha \neq \beta$; $r_{i\alpha}$ and $p_{i\beta}$ are the coordinate in the $\alpha$-direction and momentum in the $\beta$-direction of the $i$-th ion, respectively.
Thereby, the unit of $\eta$ is $\mathrm{N\cdot s/m^2}=\mathrm{Pa} \cdot \mathrm{s}$.
For simplicity, we set $\alpha=x$, $\beta=y$, and $\mathcal{L}_{\alpha\beta}(0)=0$. Hence,
\begin{equation}
(\mathcal{L}_{\alpha \beta}(t))^2
%= (\frac{1}{V} \Sigma_i r_{i\alpha}p_{i\beta})^2
= \frac{m^2}{V^2} \sum\limits_i (x_i v_{iy})^2
= \frac{m^2}{V^2} N \langle (x v_y)^2 \rangle .
\label{eq:Lab}
\end{equation}
Here, $m$ is the mass of an ion, $N$ is the total number of ions, $x$ is the coordinate in $x$-direction at time $t$, and $v_y$ is the velocity in the $y$-direction at time $t$. Obviously, due to the diffusion effect, the term that contains $x$ increases with the increase of $t$. Therefore, the quantity $\mathcal{L}_{\alpha\beta}$ is a function of $t$, and it is proportional to $t$ which will be derived later. This indicates that the factor $1/t$ in Eq. (\ref{eq:definition}) is compensated, thus the viscosity $\eta$ is independent of $t$. We can calculate the quantity $(xv_y )^2$ using the random-walk assumption (Assumption 1), i.e.,
\begin{equation}
(x v_y)^2=[(\sum\limits_{j=1}^{N_s} \Delta x_j) (\sum\limits_{j=1}^{N_s} \Delta v_{jy})]^2.
\label{eq:xvy1}
\end{equation}
Here, $\Delta x_j$ and $\Delta v_{jy}$ are the changes in $x$ and $v_y$ of the $j$-th step of an ion, respectively, and $N_s \approx \nu t$ is the total step within time $t$, where $\nu =v/\lambda_m$ is the collision frequency, and $\lambda_m$ is the mean free path of the ions. Assumption 1 permits the cross term in Eq. (\ref{eq:xvy1}) to be neglected when computing the ensemble average. Then, we obtain:
\begin{equation}
\langle (x v_y)^2 \rangle
= N_s \langle (\Delta x)^2 \rangle \langle (\Delta v_y)^2 \rangle
= N_s \frac{2}{3} \lambda_m^2 \langle (\Delta v_y)^2 \rangle.
\label{eq:xvy2}
\end{equation}
Here, we use the relation $\langle (\Delta x)^2 \rangle=\frac{2}{3} \lambda_m^2$. Thus far, the problem has been reduced to the calculation of $\langle (\Delta v_y )^2 \rangle$.

\begin{figure}[t]
\includegraphics[width=0.42\textwidth]{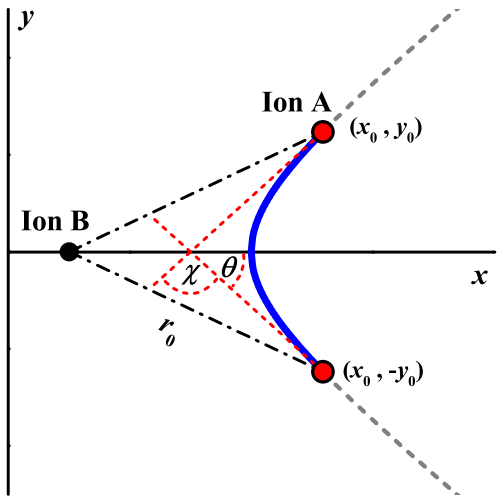}% Here is how to import EPS art
\caption{\label{fig:scheme} Hyperbolic track (blue curve) of an ion in the RWSP-VM. $\chi$ is the angle variation of the ion velocity during a ``collision''. $\theta$ is the angle between the initial ion velocity and the $x$-axis.}
\end{figure}

Assumption 2 allows the trajectory of an ion to be a hyperbolic curve when the ion-pair distance is below the cutoff distance $r_0$, as shown in Fig. \ref{fig:scheme}. Ion A moves toward ion B. The hyperbolic equation is:
\begin{equation}
 \frac{x^2}{a^2} - \frac{y^2}{b^2} = 1 \ (x>0,\ c=\sqrt{a^2+b^2}).
\label{eq:hyperbola}
\end{equation}
The parameter $a$ is equal to $\frac{q^2}{3 k_B T + 2 q^2/r_0}$, and $b$ is an independent variable. Here, $q^2=\frac{(\overline{Z}e)^2}{4 \pi \varepsilon_0}$, $\overline{Z}=\frac{1}{n} \sum\limits_{j=0}^{Z_{nc}}\ j n_j$ is the average ionization, $Z_{nc}$ is the nuclear charge, and $n_j$ is the number density of the $j$-th ionized ion. The coordinates $(x_0,y_0)$ are given as ($x_0=\frac{a}{c} (r_0-a),\ y_0=\frac{b}{c} \sqrt{(r_0-a)^2-c^2}$).
Here, the maximum value of $b$ is $b_m=\sqrt{r_0^2-2r_0 a}$, $\tan \theta=\frac{x_0 b^2}{y_0 a^2}$, and the angular variation in the ion velocity is $\chi=\pi-2\theta$. Therefore, the quantity $\langle (\Delta v_y )^2 \rangle$ can be written as
\begin{equation}
\langle (\Delta v_y )^2 \rangle
= \int_0^{b_m} \frac{1}{\nu} 2\pi b v n (v \sin \chi)^2 \mathrm{d}b
= \frac{2\pi v^3 n}{\nu} I(T),
\label{eq:dy2}
\end{equation}
where we define $I(T)=\int_0^{b_m} b (\sin\chi)^2 \mathrm{d}b$, and $n$ is the number density of the ions. Using the relations $\nu=\sqrt{2}\pi n d^2 v$ and $m v^2=3 k_B T$, the shear viscosity $\eta$ can be expressed as follows:
\begin{equation}
\eta
= \frac {\sqrt{3 m k_B T}} {\pi d^4} I(T).
\label{eq:eta}
\end{equation}
Here, $d$ is the collision diameter. For physical considerations, the parameters are set as $r_0=d$ and $d=\lambda_D$. Here, $\lambda_D \equiv \sqrt{\frac{\varepsilon_0 k_B T}{n_e e^2 (z^*+1)}}$ is the Debye length, $n_e$ is the number density of the electrons, and $z^* \equiv \overline{Z^2}/\overline{Z}$. For metals, $\overline{Z^2}=\frac{1}{n} \sum\limits_{j=0}^{Z_{nc}}\ j^2 n_j$.
Fortunately, the integration $I(T)$ has an analytical expression, written as:
\begin{equation}
I(T)=2 r_0^2 \cdot \frac{K \left [ 2(1-K) + (1+K)\mathrm{ln}K \right ]}{(1+\sqrt{K})^2 (1-K)^2} ,
\label{eq:IT}
\end{equation}
where $K=(\frac{r_0-a}{a})^2$. Finally, combining Eq. (\ref{eq:eta}--\ref{eq:IT}), we obtain an analytical expression for the viscosity.

The viscosities of metal elements in the warm dense state are calculated rapidly using Eq. (\ref{eq:eta}). Besides, in the RWSP-VM, we only require a few known and/or easily available input parameters for the ions (atom mass, nuclear charge, ionization, etc.) and several physical constants (vacuum permittivity, Boltzmann constant, and elementary charge), which is different from the case of other models.
Particularly, the most considerable difference between our model and the SMT model is Assumption 2, which leads to completely different results. The viscosity values obtained from the two models will be presented in Sec. \ref{sec:Al}.
In the remainder of this work, the coupling parameter is defined as $\Gamma=q^2/(a_{ws}k_B T)$, and the screening parameter in the SMT model is defined as $\kappa=a_{ws}/\lambda_e$, where $\lambda_e$ is the electron screening length (further details are provided in Ref. \mycite{transportMurillo}).

Details of the LMD and CMD simulations can be found in Refs. \mycite{methodsHou2} and \mycite{methodsHou}. In this work, LMD and CMD simulations are conducted to calculate the viscosities of Fe and Be, while the viscosities of Al and U are obtained from Refs. \mycite{methodsHou} and \mycite{dataU}.

\section{\label{sec:Results}Results and Discussion}
%\textbf{\textit{Average ionization}}.---
The viscosities of metal elements with low to high atomic number ($Z$) (Be, Al, Fe, and U) are calculated by employing the proposed analytical expression of RWSP-VM. One of the most significant parameters is the ionization of the metals, which affects the accuracy of the viscosity calculation. Hence, we first estimate the ionization degree using different methods, then evaluate the effect on viscosity from different average ionizations, and use the RWSP-VM to calculate the viscosities of the four different metals. Accurate comparisons and a detailed discussion are then provided.
\subsection{\label{sec:ionization}Average Ionization}
There are several methods to calculate $\overline{Z}$, such as the Saha model, Thomas--Fermi (TF) model \cite{resultsThomas}$^,$\cite{TFfit}, density functional theory molecular dynamics (DFT-MD) \cite{PIMC, DFT}, path integral Monte Carlo (PIMC) approach \cite{PIMC}, AAHNC approximation \cite{RN21}, and the Hartree--Fock--Slater (HFS) model \cite{HFS}.
The latter four methods which require the most sophisticated modelling describe the ionization and recombination, and the $\overline{Z}$ values obtained from these methods should be accurate.
%However, the TF and Saha models are used in the RWSP-VM because of their convenience. Thus, we need to evaluate the validity of the TF and Saha models.
%\textcolor{red}{It is worthy noticing that the TF theory is a theory that calculates the atomic structure, and the TF model mentioned in this work refers to the ionization state (or $\overline{Z}$) calculated from the TF theory which is defined as Eq. (51) in Ref. \mycite{TFfit}, where R. M. More determined the free electron density from the boundary density. Furthermore, R. M. More provided an approximate fit to Eq. (51) to calculate $\overline{Z}$, which is used in our model.} \textcolor{blue}{Although the TF model is not well-suited for ionization determination as illustrated in Ref. \mycite{TFFromy}, i.e., other methods are more accurate than the TF model, the following comparisons will show that the TF model is accurate in general at the conditions that we concern in this work.}

Here, we calculate $\overline{Z}$ by employing the TF (for Al, Fe, U, and Be), HFS (for Al, Fe, U, and Be), AAHNC (for Fe and Be), and Saha (for Al and Be) models. The TF model which is used in this work refers to the ionization state (or $\overline{Z}$) calculated from the TF model which is defined as Eq. (51) in Ref. \mycite{TFfit}, where More provided an approximate fit to Eq. (51) to calculate $\overline{Z}$.
We use the lowering ionization approximation reported in Ref. \mycite{RN22} in the Saha model.
\begin{figure}[t]
\includegraphics[width=0.42\textwidth]{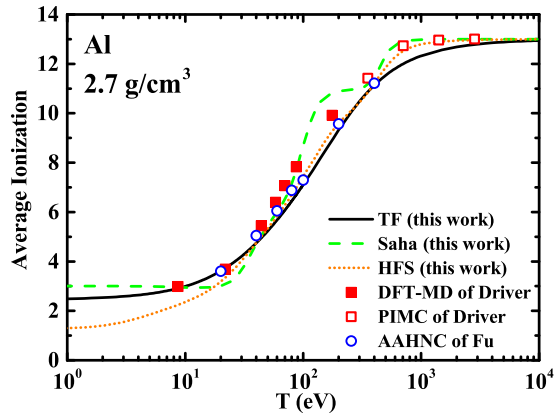}% Here is how to import EPS art
\caption{\label{fig:ZAl} Average ionization $\overline{Z}$ of Al at a density of 2.7 $\mathrm{g/cm^3}$. The black solid, green dashed, and orange dotted lines represent the results of the TF, Saha, and HFS models, respectively. The red filled and open squares indicate the DFT-MD and PIMC results reported by Driver et al. \cite{PIMC}, respectively. The blue open circles represent the AAHNC results reported by Fu et al. in a previous work \cite{RN21}.}
\end{figure}

By comprehensively analyzing the ionization obtained from the different methods, we decided to use the TF model to calculate $\overline{Z}$ in our RWSP-VM, unless otherwise specified, according to the analysis below. Fig. \ref{fig:ZAl} shows the calculated $\overline{Z}$ for the different methods, taking Al as an example.
In general, $\overline{Z}$ increases as the temperature increases. The different methods yield discrepant results in the warm dense regime.
Firstly, we compare $\overline{Z}$ obtained from the Saha and TF models with that derived by Driver \cite{PIMC} for $T<1000$ eV. The former two models deviate from the latter to within 16.6$\%$ (Saha--Driver) and 14.4$\%$ (TF--Driver).
Similarly, regarding the AAHNC model, the deviations in $\overline{Z}$ are within 16.4$\%$ (Saha--AAHNC) and 6.3$\%$ (TF--AAHNC).
Moreover, the TF results are in better agreement with the HFS results than the Saha results in the region from around 20 to 500 eV. In general, the $\overline{Z}$ value obtained from the TF model is more accurate than that obtained from the Saha model for $T< 1000$ eV.
The differences among these methods may be caused by several reasons, as noted by Driver et al. \cite{PIMC}. For example, at high temperature, the hybridization of the atomic orbitals is so large that it is difficult to separate the free and bound electrons, resulting in the difficulty to define $\overline{Z}$ rigorously. Therefore, different methods may give rise to discrepancies in the value of $\overline{Z}$ due to their particular $\overline{Z}$ definitions.
%According to the above analysis, the TF model is better than the Saha model in general.

The average ionizations of Fe, U, and Be at different densities are shown in the Appendix. In the cases of Fe and U, the TF model is more accurate at high temperature than those discussed at zero Kelvin in Ref. \mycite{TFFromy}.
%On the other hand, in the case of Be, the Saha model is more accurate for a lower density, while the TF model is more accurate for a higher density. The influence of $\overline{Z}$ on the viscosity for Be calculated from the Saha and TF models will be discussed in detail later.

The viscosity coefficients from all ionizations in Fig. \ref{fig:ZAl} are calculated, and it is obvious that the TF model is usually accurate at the conditions considered in this work. In the case of Be, the Saha model is more accurate for a lower density, while the TF model is more accurate for a higher density. The influence of $\overline{Z}$ on the viscosity for Be calculated from the Saha and the TF models will be discussed in detail later. Actually, our motivation is that we develop algorithm for accurate and highly efficient calculation of plasma viscosities in the warm dense region. The accurate viscosity evaluated with ionization of the TF model does not require the sophisticated modelling and time--consuming calculations, such as DFT-MD etc.

Once $\overline{Z}$ is obtained, the viscosities can be easily calculated using Eq. (\ref{eq:eta}).

\subsection{\label{sec:Al}RWSP-VM results for Al}
\begin{figure}[t]
\includegraphics[width=0.42\textwidth]{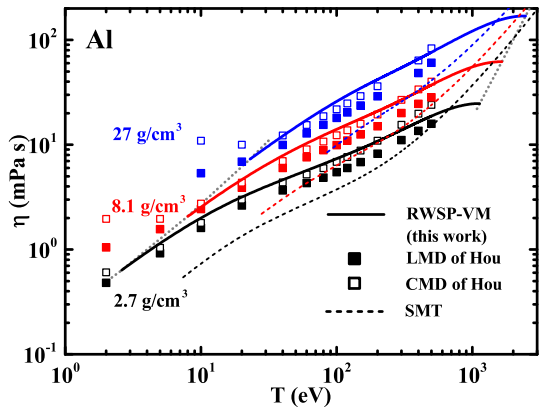}% Here is how to import EPS art
\caption{\label{fig:Al} Shear viscosity of Al. The solid lines indicate the RWSP-VM results, while the dashed lines represent the SMT results. The filled and open squares indicate the LMD and CMD simulation data reported by Hou et al. \cite{methodsHou}, respectively. The black, red, and blue colors denote the data for densities of 2.7, 8.1, and 27 $\mathrm{g/cm^3}$, respectively. The gray dotted lines represent the lower (left) and upper (right) limits of the temperature range in which the proposed model is applicable.}
\end{figure}

Al is a fundamental metal in many fields, and its viscosity is often beneficial to warm-dense-matter applications, for which the RWSP-VM might agree better with the CMD simulations than with the LMD simulations. Fig. \ref{fig:Al} shows the viscosity of Al in the temperature range from around 2 to 1000 eV and at densities of 2.7, 8.1, and 27 $\mathrm{g/cm^3}$.
The solid curves represent the RWSP-VM results. It can be seen that the viscosity increases with increasing temperature and density.

However, when the temperature is too low, the RWSP-VM is not applicable as it shows that the viscosity still increases as the temperature increases. Actually, the trend is opposite to that reported in our previous work \cite{DPvicosity}. As the Debye length is very small ($\lambda_D<0.1 a_{ws}$), the Coulomb potential considered in the RWSP-VM is not the only interatomic interaction, other interactions play more important roles. Here, $a_{ws}=(\frac{3}{4\pi n})^{(1/3)}$ is the Wigner--Seitz radius of the ion. Therefore, we use the temperature $T_{\mathrm{lower}}$ at which $\lambda_D(T_{\mathrm{lower}})=0.1 a_{ws}$ to cut off the viscosity curves, which corresponds to the gray dashed line on the left-hand side of the curves. In other words, the RWSP-VM is applicable for $T>T_{\mathrm{lower}}$.

Besides, we notice that the viscosity obtained from the RWSP-VM starts to decrease with increasing temperature as the temperature exceeds $10^3$ eV, which is in disagreement with the results reported by Ref. \mycite{methodsDaligault}. The reason for this discrepancy is that ``introducing a truncated range in the impact parameter is not equivalent to truncating the range of the Coulomb interaction'' \cite{transportMurillo}. When the temperature is sufficiently high, the Debye shielding effect is reduced, and multibody collisions become significant; thus, the quantity $\langle (\Delta v_y )^2 \rangle$ is underestimated due to Assumption 1. Here, we use the temperature $T_{\mathrm{upper}}$ at which $\lambda_D(T_{\mathrm{upper}})=0.5 a_{ws}$ to cut off the viscosity curves. As for $T_{\mathrm{lower}}$, the upper cutoff line is indicated by the gray dashed line on the right-hand side of the curves. In other words, the RWSP-VM is applicable for $T<T_{\mathrm{upper}}$.

Furthermore, the results of the SMT model are shown in the figure (the dashed curves) for comparison. Clearly, our model is considerably different from the SMT model and agrees well with both the LMD and CMD simulations. This is because the screening length plays an important role in the SMT model.
In the case of Fig. \ref{fig:Al}, for the SMT model in the warm dense regime, the screening parameter $\kappa$ is close to $\kappa=0$ according to the definition in the SMT model. In addition, we believe that the SMT model might not be applicable for $\kappa \to 0$ in the warm dense regime. The reason may be that the SMT model results agree well with the MD data for $\Gamma < 10$ at $\kappa=1$, $\Gamma < 20$ at $\kappa=2$, and $\Gamma < 100$ at $\kappa=3$. Although the case of $\kappa=0$ is not explored in this previous work \cite{transportMurillo}, it is reasonable to infer that for $\kappa=0$, the applicability range of $\Gamma$ would be considerably smaller than that of $\kappa=1$, i.e., the range considered in our model is outside  the applicability range of the SMT model. Hence, the proposed RWSP-VM is substantially different from the SMT model.
The viscosity values obtained from the RWSP-VM deviate from those obtained from the LMD and CMD simulations to within 46.4$\%$ and 21.6$\%$, respectively. This indicates that the RWSP-VM results agree better with those of the CMD simulations rather than those of the LMD simulations.

The discrepancy between our model and the SMT model is due to the fact that we use a cutoff distance to describe the shielding effect, while the SMT model uses the screening length to evaluate this effect. Our model introduces the hyperbolic curve with the help of the cutoff distance to derive the expression, i.e., Eq. (\ref{eq:eta}), and the upper limit of the integral ($b_m$) in Eq. (\ref{eq:dy2}) is obtained naturally. On the other hand, the SMT model introduces the screening length according to the Coulomb logarithm, which is also the upper limit of the integral in Eq. (5) for the SMT model (a divergence would be encountered if the upper limit tends to infinity), and should be evaluated more carefully to obtain more accurate results. In other words, in the SMT model, different screening lengths result in different viscosity values, and $\lambda_{\mathrm{eff}}$ (for the details see Ref. \mycite{transportMurillo}) is selected for $\kappa$ in the range from 1 to 3. In general, these two models differ due to the different treatment of the screening effect, and based on the comparison of the results of the two models with those of the LMD and CMD simulations, we believe that the SMT model is not accurate in the warm dense regime unless it can evaluate the screening length more appropriately.
Therefore, the SMT model is not used for the comparisons in the remainder of this work.

\subsection{\label{sec:Fe}RWSP-VM results for Fe}
\begin{figure}[tb]
\includegraphics[width=0.42\textwidth]{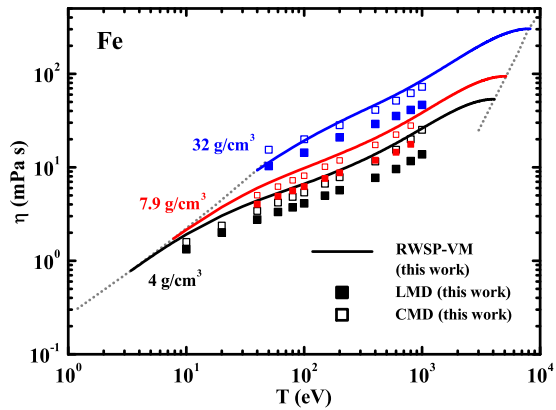}% Here is how to import EPS art
\caption{\label{fig:Fe} Shear viscosity of Fe. The solid lines show the RWSP-VM results.
The filled and open squares represent the LMD and CMD data calculated in this work, respectively.
The black, red, and blue colors denote the data for densities of 4, 7.9, and 32 $\mathrm{g/cm^3}$, respectively.
The gray dotted lines represent the lower (left) and upper (right) limits of the temperature range in which the proposed model is applicable.
}
\end{figure}

Fe is a key element when investigating the core of terrestrial planets \cite{appFe1,appFe2,appFe3}. Fig. \ref{fig:Fe} shows the viscosity of Fe. Since Fe is a medium-$Z$ element, it behaves similar to Al.
The viscosity increases with increasing temperature and density. This indicates that the RWSP-VM results are in better agreement with those of the CMD model rather than those of the LMD model. The viscosity values derived from the RWSP-VM deviate from those obtained from the CMD model to within 29.0$\%$, 21.1$\%$, and 27.2$\%$ at densities of 4, 7.9, and 32 $\mathrm{g/cm^3}$, respectively, which shows a good agreement between the two models.
Furthermore, we compare the results of our model with those of the TFMD simulations reported by Daligualt et al. \cite{methodsDaligault} as well as those of both the OFMD and short-range repulsion (SRR) simulations reported by Sun et al. \cite{dataFe} (not shown).
Especially, the values obtained from the RWSP-VM deviate from those derived from the OFMD and SRR simulations by 0.51$\%$--46.1$\%$ and 0.17$\%$--19.6$\%$, respectively. This is because the OFMD simulations based on the TF approximation are not applicable at low temperatures and densities, while the SRR simulations are more accurate as they utilize the Yukawa model and the corresponding repulsion potential. This repulsion potential is similar to the shielding Coulomb potential used in this work.
Due to the fact that the AAHNC method is more accurate than the TFMD, OFMD, and SRR simulations, the AAHNC method is here selected to calculate the viscosity and  is used for the comparisons.
Here, the lower and upper temperature limits are still valid in the case of Fe.
\subsection{\label{sec:U}RWSP-VM results for U}
\begin{figure}[t]
\includegraphics[width=0.42\textwidth]{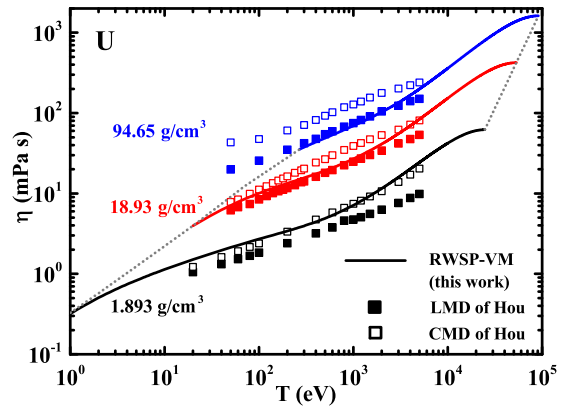}% Here is how to import EPS art
\caption{\label{fig:U} Shear viscosity of U.
The legend is the same as that in Fig. \ref{fig:Al}. The LMD and CMD data are from a previous work \cite{dataU}.
The black, red, and blue colors denote the data at densities of 1.893, 18.93, and 94.65 $\mathrm{g/cm^3}$, respectively.
}
\end{figure}

U is often used in planar and hohlraum targets in ignition experiments. Here, at higher densities, the RWSP-VM results are in better agreement with the results of the LMD model, while they are in better agreement with the results of the CMD model at lower densities. Fig. \ref{fig:U} shows the viscosity of U. The viscosity increases with increasing temperature and density. At a density of 1.893 $\mathrm{g/cm^3}$, the RWSP-VM-derived viscosity agrees better with that obtained from the CMD model rather than that obtained from the LMD model. Especially, the values obtained from the RWSP-VM deviate from those obtained from the LMD and CMD simulations by 35.9$\%$--147$\%$ and 1.0$\%$--23.4$\%$, respectively. By contrast, at densities of 18.93 and 94.65 $\mathrm{g/cm^3}$, the RWSP-VM results agree better with those of the LMD model rather than those of the CMD model except for a few data at the highest temperatures. Especially, the values obtained from the RWSP-VM deviate from those derived from the LMD and CMD models by 1.7$\%$--20.4$\%$ and 3.0$\%$--47.8$\%$, respectively (except for a few data at the highest temperatures, around 5000 eV).

This phenomenon may be explained by the fact that ``the electron--ion dynamic collisions (introduced by LMD) increase the effective collision cross section and weaken the interaction between ions'' \cite{dataU}. In detail, for high-$Z$ elements at higher densities, with increasing temperature, the charge states and ionic structures become more complex, and numerous free electrons are produced due to the pressure and thermal ionization, which influence the motion of ions, thus influencing the viscosity. This results in the fact that the nonadiabatic dynamic effects weaken the interaction between ions at higher densities. Therefore, the LMD model is more appropriate than the CMD model in this case. The LMD model corresponds to the case of the Debye shielding effect in our model. That is, for high-$Z$ elements at higher densities, the electron--ion collision effect plays an important role, which is equivalent to introducing a cutoff distance, and the interaction between ions is weakened, which is similar to the LMD case. Therefore, the results of our model agree better with those of the LMD simulations in this case.
\subsection{\label{sec:Be}RWSP-VM results for Be}
\begin{figure}[t]
\includegraphics[width=0.42\textwidth]{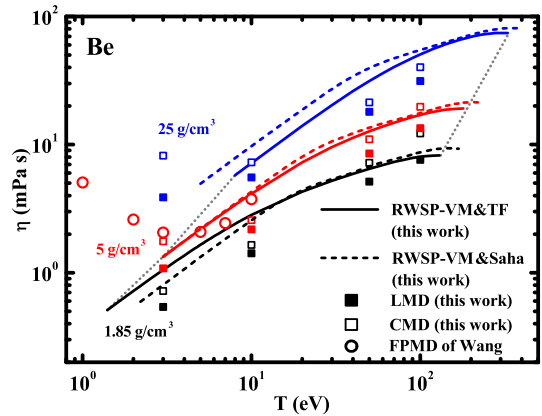}%
\caption{\label{fig:Be} Shear viscosity of Be. The solid and dashed lines show the RWSP-VM results employing the TF and Saha models, respectively. The filled and open squares represent the LMD and CMD results, respectively. The open circles indicate the FPMD data reported by Wang et al. \cite{dataBe}. The black, red, and blue colors denote the data for densities of 1.85, 5, and 25 $\mathrm{g/cm^3}$, respectively. The gray dotted lines represent the lower (left) and upper (right) limits of the temperature range in which the proposed model is applicable (only the RWSP-VM employing the TF model is depicted here for clarity).}
\end{figure}

Be plays an important role as an ablator in ICF \cite{ICFBe}. Here, the RWSP-VM results agree well with those of the CMD and FPMD simulations but not with those of the LMD simulations. Additionally, the TF model is more accurate at higher densities, while the Saha model is more accurate at lower density when estimating $\overline{Z}$. Fig. \ref{fig:Be} shows the shear viscosity of Be. Wang et al. \cite{dataBe} employed the FPMD model at a density of 5 $\mathrm{g/cm^3}$, while we employ the LMD and CMD models at densities of 1.85, 5.0, and 25 $\mathrm{g/cm^3}$. For the FPMD (5 $\mathrm{g/cm^3}$) and CMD (25 $\mathrm{g/cm^3}$) data, the viscosity first decreases and then increases with increasing temperature. This is because for these two cases, the temperature ranges of applicability are from the low-temperature regime to the warm regime, where there is a competition between the kinetic and potential components of the energy of the particles. Both components contribute to the viscosity. When the temperature is low, the kinetic energy component is negligible compared with the potential energy one, resulting in a decrease in the viscosity due to the decrease in the potential energy with the increase in temperature. As the temperature increases, the kinetic energy component is no longer negligible, resulting in an increase in the viscosity due to the increase in the kinetic energy component with increasing temperature. This phenomenon has been illustrated in Ref. \mycite{modelsPostovalov} and our previous work \cite{DPvicosity}. In the warm dense regime, the viscosity increases with increasing temperature.

Here, we consider both the TF and Saha models to calculate the average ionization (shown in Fig. \ref{fig:ZBe}) for comparison. We notice that the viscosity obtained from the RWSP-VM employing the TF model is not in good agreement with those obtained from the CMD and LMD simulations at 1.85 $\mathrm{g/cm^3}$ and 3 eV, and the viscosity obtained from the RWSP-VM employing the Saha model does also not agree well with them at 25 $\mathrm{g/cm^3}$. This is because for low (1.85 $\mathrm{g/cm^3}$) and high (25 $\mathrm{g/cm^3}$) densities, the $\overline{Z}$ values of the TF and Saha models differ substantially, which influences the RWSP-VM viscosity result. Especially, at lower temperatures (3 eV), the $\overline{Z}$ value of the TF model at 1.85 $\mathrm{g/cm^3}$ and that of the Saha model at 25 $\mathrm{g/cm^3}$ are underestimated, resulting in the inaccuracy of the RWSP-VM when employing the TF (1.85 $\mathrm{g/cm^3}$) and Saha (25 $\mathrm{g/cm^3}$) models in the corresponding cases. At a density of 5 $\mathrm{g/cm^3}$, there is a less pronounced difference in the $\overline{Z}$ value between the TF and Saha models, and the two viscosity results are consistent with each other. Therefore, the Saha model is more suitable at lower densities (1.85 $\mathrm{g/cm^3}$), while the TF model is more appropriate at higher densities (5 and 25 $\mathrm{g/cm^3}$). This corresponding (the Saha model for 1.85 $\mathrm{g/cm^3}$ and the TF model for 5 and 25 $\mathrm{g/cm^3}$) is used for the comparisons below. Especially, the viscosity obtained from the RWSP-VM deviates from that derived from the CMD simulations by 6.3$\%$--28.9$\%$ at densities of 1.85 and 5 $\mathrm{g/cm^3}$ except for the 10-eV temperature case. The deviation between the RWSP-VM and CMD results is 1.2$\%$--50.0$\%$ at 25 $\mathrm{g/cm^3}$, while the deviation between the RWSP-VM and FPMD results is 4.7$\%$--35.2$\%$. In general, the viscosity obtained from the RWSP-VM agrees well with those obtained from the CMD and FPMD simulations under the same conditions. As Be is a low-$Z$ element, the ionization is sufficiently low  (less than 4), which weakens the influence of electrons. This is the reason why the CMD model is here preferred over the LMD model for the comparisons (which is the same reason as for the Al case)

\subsection{\label{sec:comparison}Comparison of the RWSP-VM with the YVM and OCP models}
\begin{figure}[t]
\includegraphics[width=0.42\textwidth]{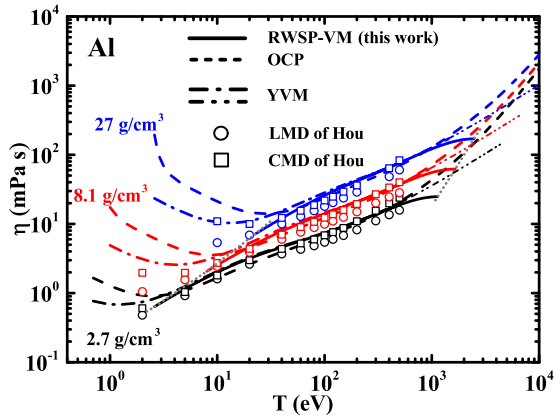}%
\caption{\label{fig:ComAl} Shear viscosity of Al. The solid, dashed, and dash--dotted lines show the RWSP-VM, OCP, and YVM results, respectively. The dash--dot--dot lines represent the extension of the YVM range for $1/3 \le \Gamma \le 2$. The open circles and squares indicate the LMD and CMD results reported by Ref. \cite{methodsHou}, respectively. The black, red, and blue colors represent the data stand for densities of 2.7, 8.1, and 27 $\mathrm{g/cm^3}$, respectively. The gray dotted lines represent the lower (left) and upper (right) limits of the temperature range in which the proposed model is applicable.}
\end{figure}

In the introduction, we  mentioned two viscosity models, i.e., the YVM and OCP models.
The models used in this section refer to the physical models reported in Refs. \mycite{modelsMurillo} and \mycite{modelsDaligault} and their corresponding parameters, respectively. The YVM parameters were obtained by fitting the MD data taken from Ref. \mycite{modelsSaigo}.
%Usually, the coupling parameter $\Gamma=q^2/(a_{ws}k_B T)$ is introduced as a dimensionless quantity to reveal the coupling strength of plasmas.
On the other hand, the OCP model is applicable for one-component plasmas from the weakly coupled regime (high temperatures) to the moderately (warm dense matter) and strongly (liquid) coupled regimes. The OCP parameters were derived from the fitting of the equilibrium MD data.
Here, we take Al as an example to compare our model with these two models, as shown in Fig. \ref{fig:ComAl}.
For both the YVM and OCP models, the viscosity first decreases and then increases as the temperature increases. The reason behind this phenomenon was explained in Sec. \ref{sec:Be}. In the warm dense state (middle range), the three models agree well with each other. Particularly, our model is in a slightly better agreement with the YVM model than the OCP model. In the low-temperature range, the results of the YVM model are in better agreement with the CMD results than the OCP results. This is because the OCP expression used here employs $\kappa=0$, which is not a good approximation in low- and warm-temperature ranges, while the YVM expression employs $\kappa$ from 0.1 to 3. In the high-temperature range, the OCP model is reliable as, in the weakly coupled, it agrees well with the Landau--Spitzer prediction, as reported by \cite{modelsDaligault}. In the case of the YVM model, we extend the comparison range from $\Gamma \ge 2$ to $\Gamma \ge 1/3$, as shown by the dash--dot--dot lines.
It is clear that the difference between the YVM and OCP results becomes more pronounced with increasing temperature owing to the different applicability range of each model. This is consistent with Murillo's speculation, according to which the YVM model underpredicts the viscosity at a weak coupling.

The comparisons for the other three elements are shown in the Appendix. In general, the results indicate that in the range of temperatures in which the RWSP-VM is applicable, the results of the three models are consistent with each other.
The YVM model is more suitable at lower temperatures, while the OCP model is more appropriate at higher temperatures. None of the models is applicable across the entire temperature range.

\subsection{\label{sec:app}Implementation and application of the RWSP-VM}
The main characteristics of the RWSP-VM are its universality, accuracy, and high efficiency. The reasons for these beneficial properties are summarized below.
Firstly, this model is based on the random-walk, ion gas, and shielding-potential assumptions, which physically describe warm dense metals. Accordingly, the different types of metal ions behave similarly in the warm dense state.
Secondly, the shielding potential is treated as the Coulomb potential by employing a cutoff distance described by the Debye length, which is a good approximation for warm dense metals.
Thirdly, ``binary collisions'' are the dominant event of the dynamic processes in the warm dense regime, and this assumption makes it simple to obtain a simple expression.
This model is applicable to elemental metals with low to high Z, temperatures from several eV to hundreds or even thousands of eV, and densities from 0.1 to 10 times times the normal density (the density at room temperature and 1 standard atmosphere).

\section{\label{sec:Conclusions}Conclusions}
%\textbf{\textit{Conclusion}}.---
Taking into account the statistics of random-walk ions and the Debye shielding effect, we developed a new model, called the RWSP-VM, which is applicable to arbitrary elements. Based on the RWSP-VM, we estimated the shear viscosities of a series of metal elements in the warm dense state. The comparisons of the RWSP-VM results with the CMD and LMD results of this and previous works validate the accuracy of the proposed model. In general, methods based on MD simulations, such as FPMD simulations, require several days to simulate only one state point, while our model provides the full results within a second. The RWSP-VM ability to calculate the shear viscosity of warm dense metals will make this model applicable to numerous areas; furthermore, it is envisaged that the RWSP-VM will become important in the field of warm dense matter. Moreover, we compared our model with the OCP and YVM models. The three models were found to be in good agreement with each other in the region in which the RWSP-VM is applicable. However, it is remarkable that these models behave differently especially in the lower- and higher-temperature ranges, which indicates that developing a viscosity model applicable to a wide temperature range remains an open problem to be solved in the future.

%~\\ \indent% add an empty line
\section*{\label{sec:Acknowledgment}Acknowledgment}
We thank Shuaichuang Wang and Cong Wang for their helpful discussions. This work was financially supported by the Science Challenge Project (Grant No. TZ2016001) and the Foundation of LCP. Hou was supported by the Science Challenge Project (Grant No. TZ2018005) and National Natural Science Foundation of China (Grant No. 11974424).

\section*{\label{sec:Conflicts}Author Declarations}
The authors have no conflicts to disclose.

\section*{\label{sec:Data}Data availability}
The data that support the findings of this study are available from the corresponding author upon reasonable request. We also provide the codes for the proposed model, which can be downloaded from the Supplementary Materials \cite{SM}. See the Supplementary Materials for more details involved in this study. [URL: http://link.aps.org/supplemental/10.1103/PhysRevE.106.014142]
%or from the following GitHub repository: \href{ https://github.com/cyqphy/RWSP-VM}{\textcolor{blue}{https://github.com/cyqphy/RWSP-VM}}.

%\pagebreak
\newpage
\begin{appendix}
\renewcommand\thefigure{A\arabic{figure}}
\section*{\label{sec:Appendix}Appendix}
\subsection*{\label{sec:AppendixA} A.  The average ionizations of Fe, U, and Be}
\setcounter{figure}{0}
The average ionizations of Fe, U, and Be at different densities are shown in Fig. \ref{fig:ZFe}, \ref{fig:ZU}, and \ref{fig:ZBe}, respectively. The difference in the viscosity values obtained using different methods is provided in the main text, while the difference in the $\overline{Z}$ values is analyzed below. From Fig. \ref{fig:ZFe}, it can be seen that the TF, HFS, and AAHNC results agree well with each other except at lower temperatures. Fig. \ref{fig:ZU} reveals that the relative differences in the results obtained from the TF, HFS, and AANHC models are small due to the large atomic number of U.
As a result, the TF model is applicable to calculate the average ionizations of Fe, U, and Al. However, the TF model is not applicable in the case of Be.
From Fig. \ref{fig:ZBe} it can be seen that at a density of 1.85 $\mathrm{g/cm^3}$, compared with the HFS and AAHNC models, the TF model behaves better than the Saha model, especially at lower temperatures ($T<30$ eV). On the other hand, at 5 $\mathrm{g/cm^3}$, the TF and Saha models agree well with the HFS and AAHNC models except for at lower temperatures. Moreover, at 25 $\mathrm{g/cm^3}$, the Saha model is in better agreement with the HFS model than the TF model, while none of the three models is consistent with the AAHNC model. The relatively large errors may originate from the low $Z$ of Be.
Hence, different methods to calculate $\overline{Z}$ should be employed for different Be densities, i.e., the TF model for lower densities and the Saha model for higher densities. This has been implemented in this study to calculate the viscosity of Be, as explained in the main text.
\begin{figure}[b]
\includegraphics[width=0.42\textwidth]{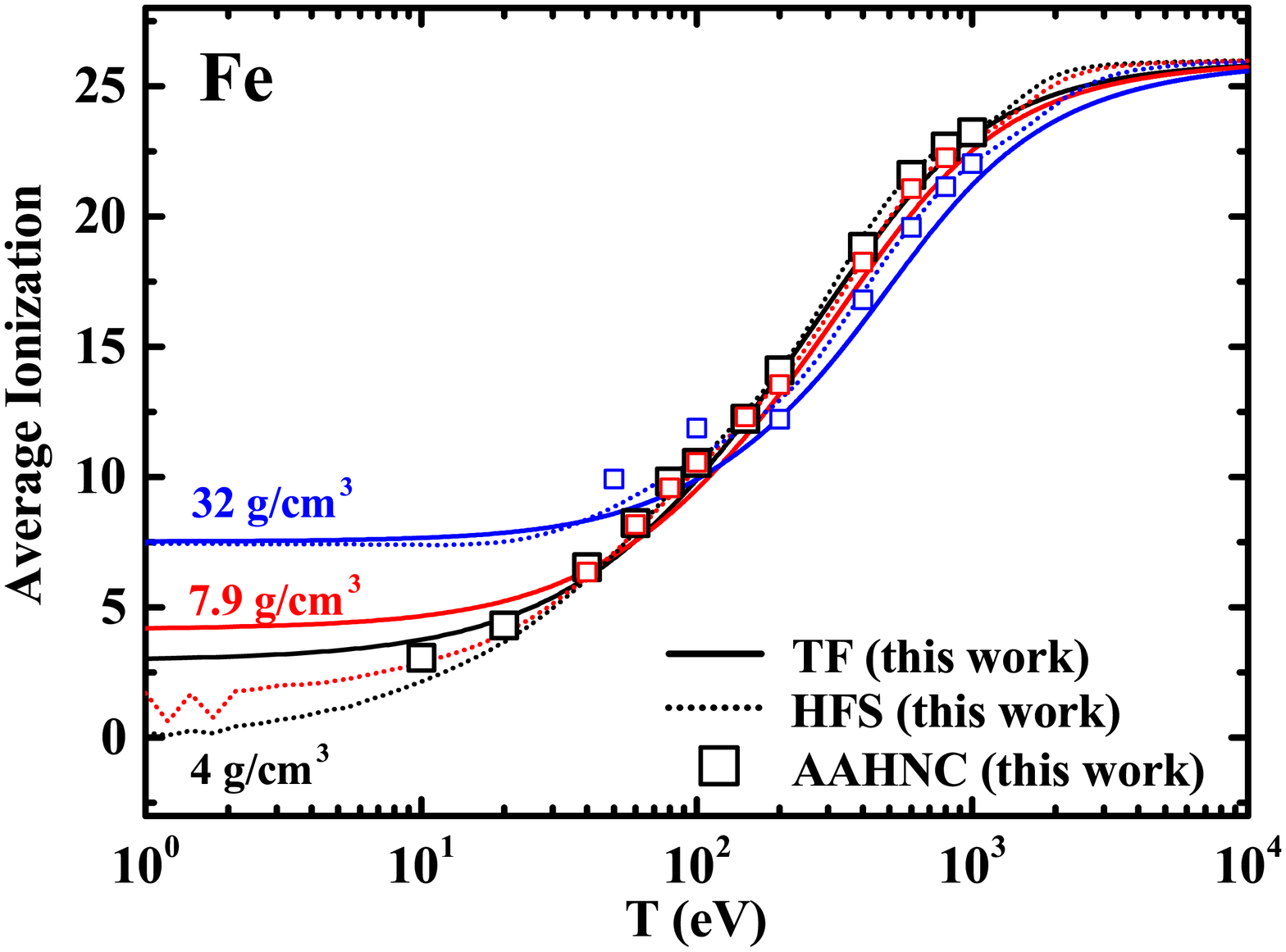}% Here is how to import EPS art
\caption{\label{fig:ZFe} Average ionization $\overline{Z}$ of Fe at densities of 4 (black), 7.9 (red), and 32 $\mathrm{g/cm^3}$ (blue). The solid and dotted lines represent the results of the TF and HFS models, respectively. The open squares represent the results of the AAHNC model obtained in this work.}
\end{figure}
\begin{figure}[t]
\includegraphics[width=0.42\textwidth]{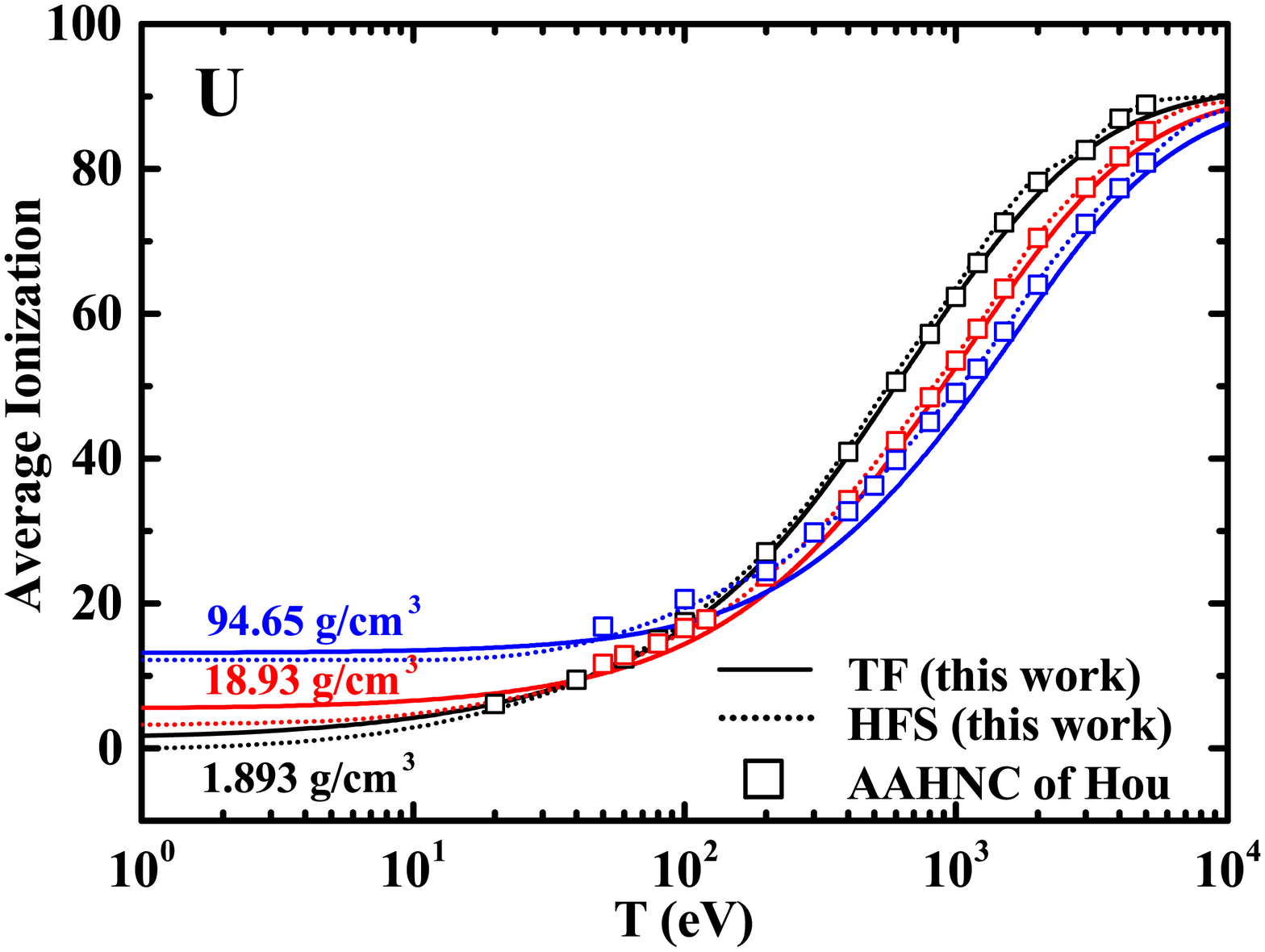}% Here is how to import EPS art
\caption{\label{fig:ZU} Average ionization $\overline{Z}$ of U at densities of 1.893 (black), 18.93 (red), and 94.65 $\mathrm{g/cm^3}$ (blue).
The AAHNC data are taken from a previous work \cite{dataU}.
The legends are the same as those in Fig. \ref{fig:ZFe}.
}
\end{figure}
\begin{figure}[t]
\includegraphics[width=0.42\textwidth]{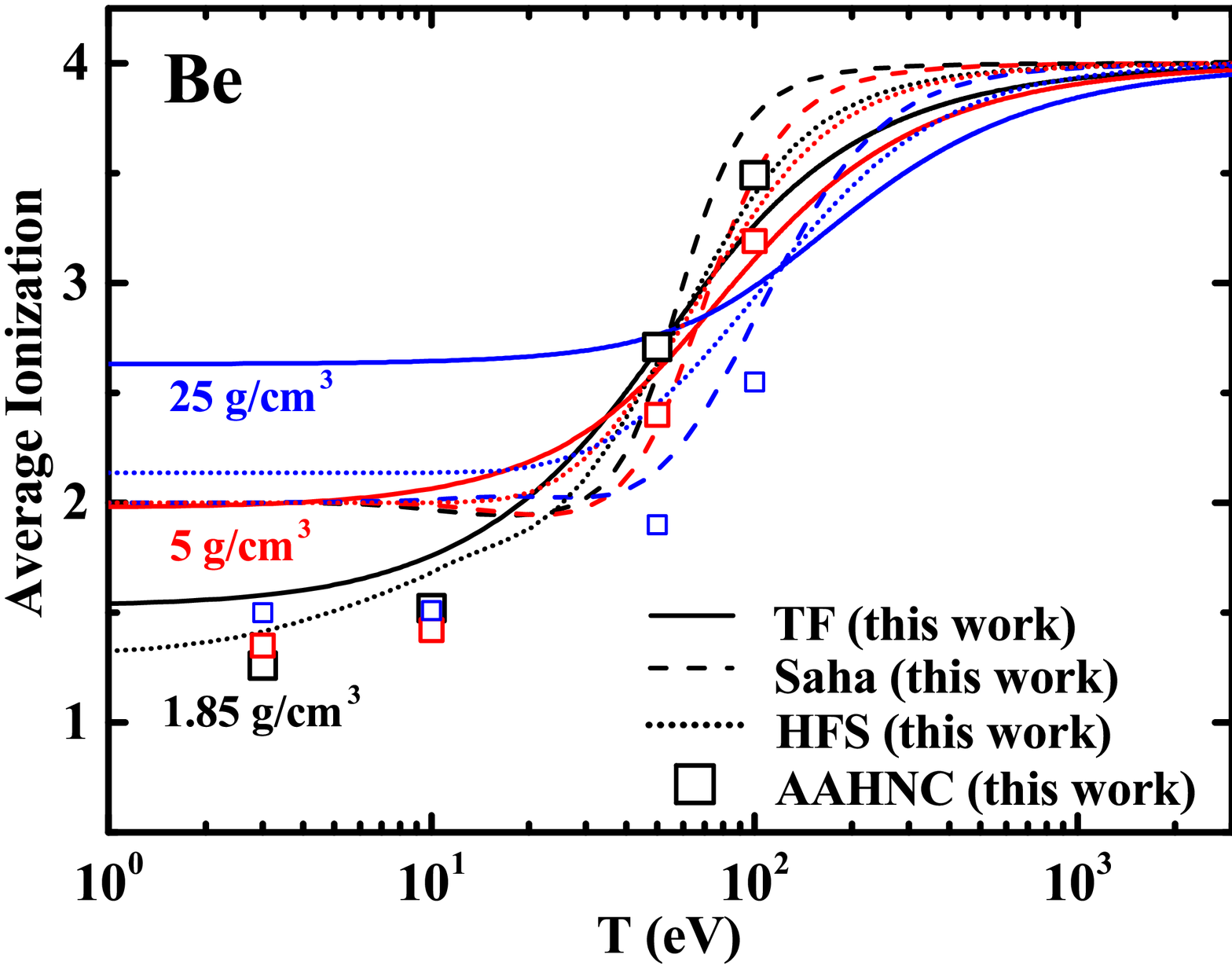}% Here is how to import EPS art
\caption{\label{fig:ZBe} Average ionization $\overline{Z}$ of Be at densities of 1.85 (black), 5 (red), and 25 $\mathrm{g/cm^3}$ (blue). The dashed line indicates the results of the Saha, and other legends are the same as those in Fig. \ref{fig:ZFe}.}
\end{figure}

\subsection*{\label{sec:AppendixB} B.  The comparison of viscosities from different models for Fe, U, and Be}
\begin{figure}[htb]
\includegraphics[width=0.42\textwidth]{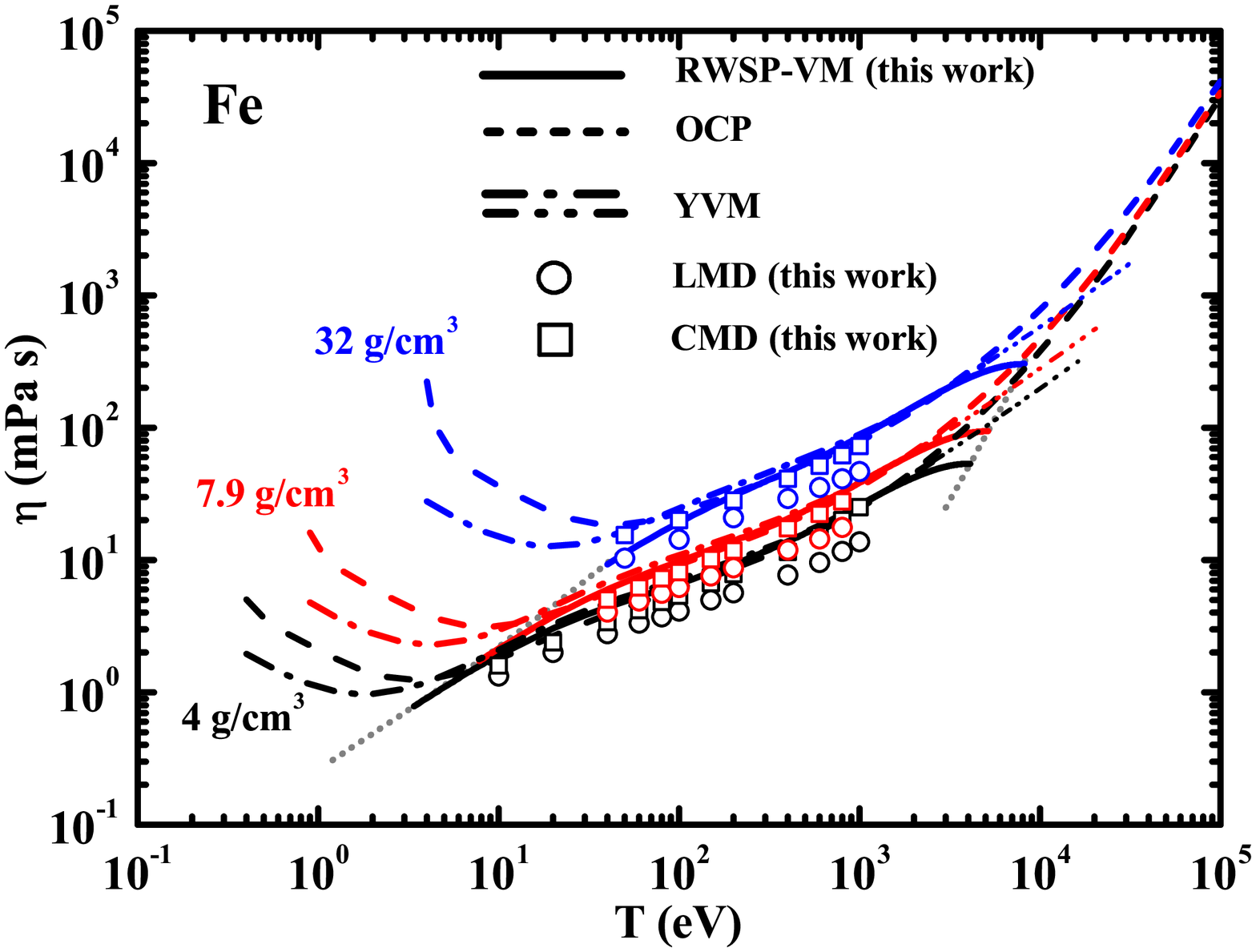}%
\caption{\label{fig:ComFe} Shear viscosity of Fe. The solid, dashed, and dash-dotted lines represent the RWSP-VM, OCP, and YVM results, respectively. The dash--dot--dot line indicates the extension of the YVM range ($1/3 \le \Gamma \le 2$). The open circles and squares indicate the LMD and CMD results obtained in this work, respectively. The black, red, and blue colors denote the data for densities of 4, 7.9, and 32 $\mathrm{g/cm^3}$, respectively. The gray dotted lines represent the lower (left) and upper (right) limits of the temperature range in which the proposed model is applicable.}
\end{figure}

\begin{figure}[htb]
\includegraphics[width=0.42\textwidth]{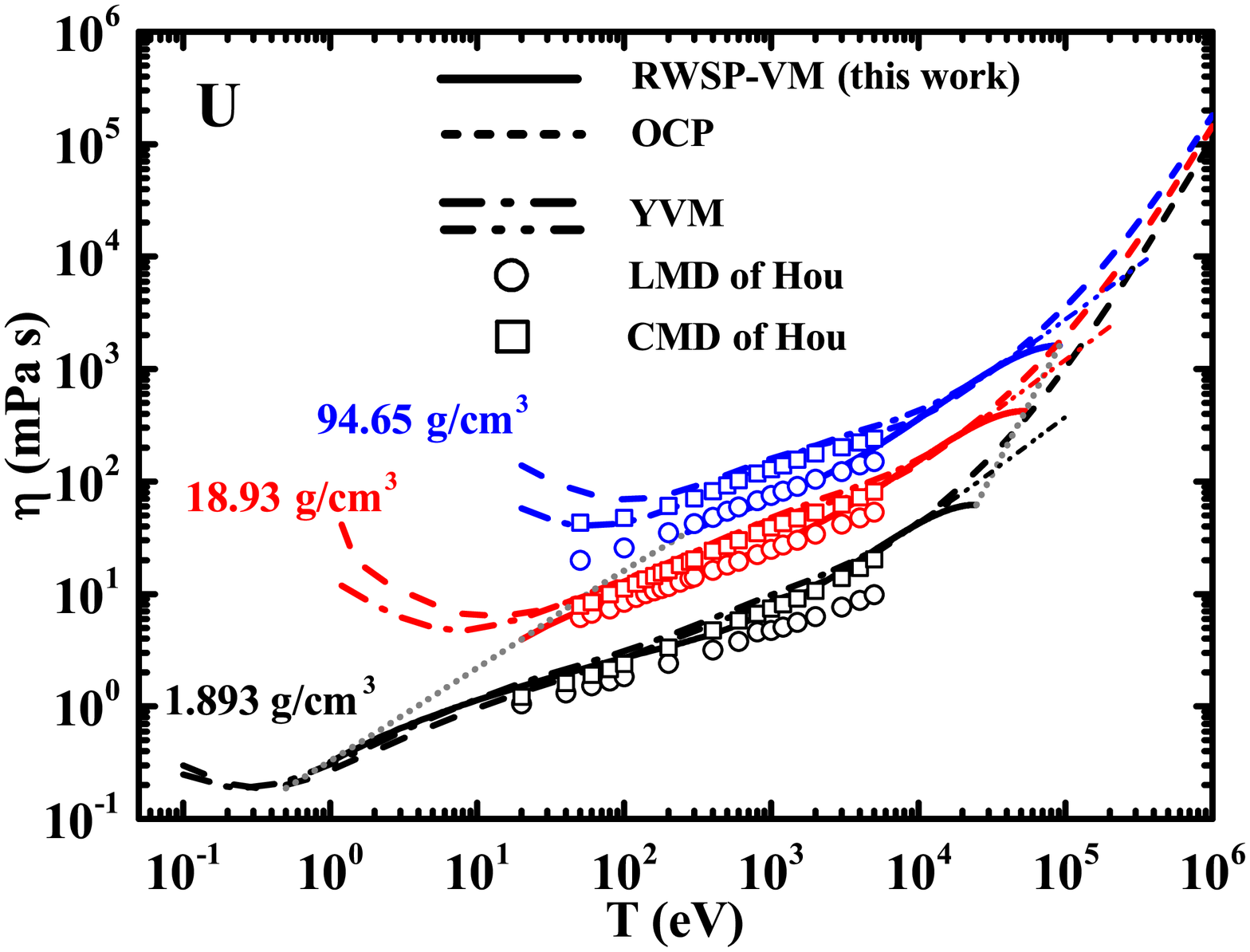}%
\caption{\label{fig:ComU} Shear viscosity of U.
%The solid, dashed, and dash-dotted lines correspond to the RWSP-VM, OCP, and YVM results, respectively. The dash-dot-dot lines represent the extension of the YVM range ($1/3 \le \Gamma \le 2$). The open circles and squares indicate the LMD and CMD results obtained in Ref. \cite{dataU}, respectively.
The black, red, and blue colors represent the data for densities of 1.893, 18.93, and 94.65 $\mathrm{g/cm^3}$, respectively.
The legends are the same as those in Fig. \ref{fig:ComFe}.}
%The gray dotted lines represent the lower (left) and upper (right) limits of the temperature range in which the proposed model is applicable.}
\end{figure}

\begin{figure}[htb]
\includegraphics[width=0.42\textwidth]{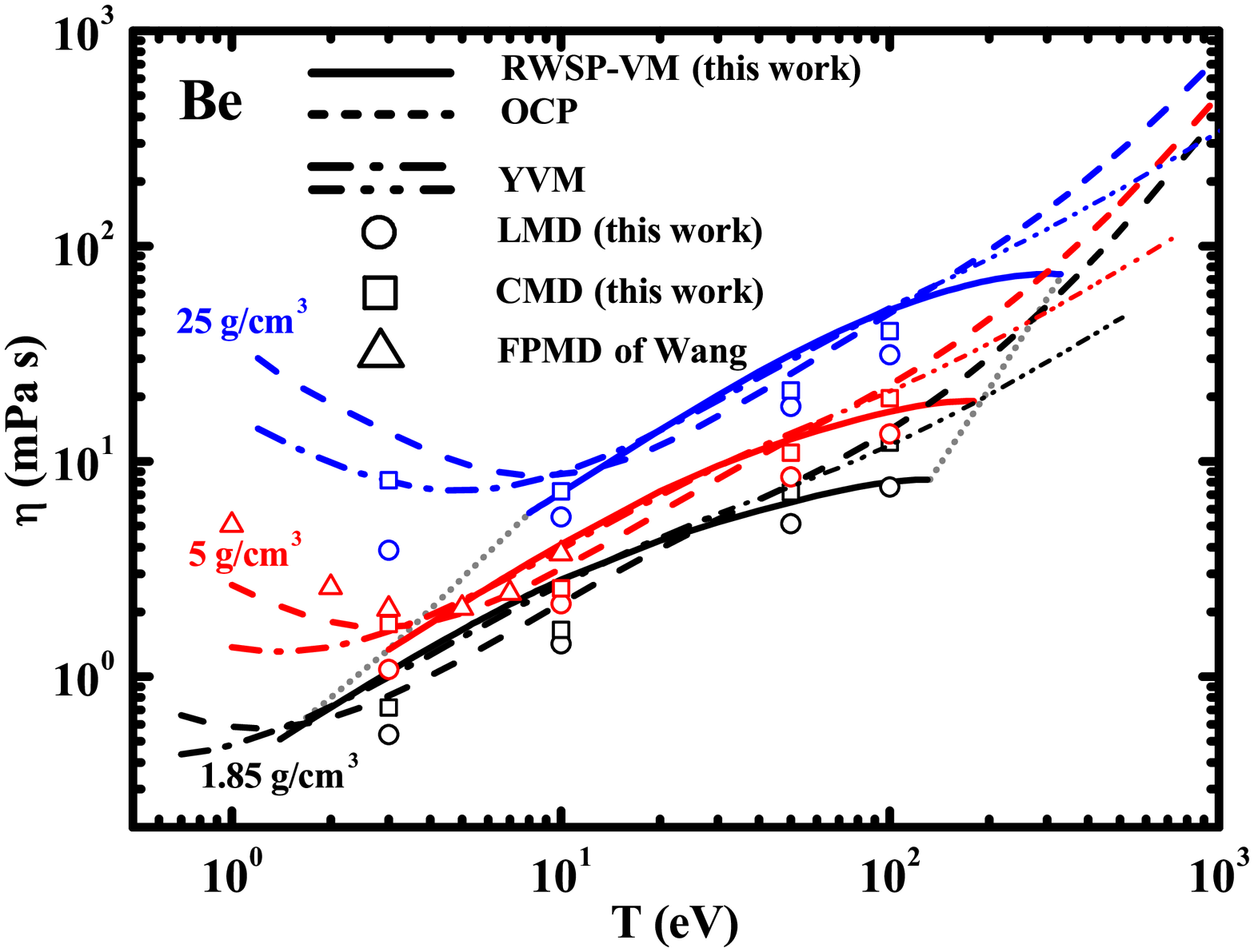}%
\caption{\label{fig:ComBe} Shear viscosity of Be.
%The solid, dashed, and dash-dotted lines indicate the RWSP-VM, OCP, and YVM results, respectively. The dash-dot-dot lines represent the extension range of the YVM model ($1/3 \le \Gamma \le 2$).
The open triangles stand for the FPMD results obtained in Ref. \cite{dataBe}.
The black, red, and blue colors denote the data for densities of 1.85, 5, and 25 $\mathrm{g/cm^3}$, respectively.
Other legends are the same as those in Fig. \ref{fig:ComFe}.}
%The gray dotted lines represent the lower (left) and upper (right) limits of the temperature range in which the proposed model is applicable.}
\end{figure}

Figs. \ref{fig:ComFe}, \ref{fig:ComU}, and \ref{fig:ComBe} show the comparisons of the models and MD results.
As the temperature increases, the viscosities decrease in low--temperature range, then increase in warm dense range, and continue to increase in high--temperature range.
The conclusions are very similar to those of Fig. \ref{fig:ComAl}.
These three models behave similarly and agree well with the CMD and LMD simulations in the warm dense range.

\end{appendix}

%~\\
\pagebreak
%\newpage
\section*{\label{sec:Ref}References}
\bibliography{viscosity_Cheng}
%\balance
%\bibColoredItems{red}{TFFromy, TFfit, SM}
%\bibliography{viscosity_Cheng}
%merlin.mbs aipnum4-1.bst 2010-07-25 4.21a (PWD, AO, DPC) hacked
%Control: key (0)
%Control: author (8) initials jnrlst
%Control: editor formatted (1) identically to author
%Control: production of article title (0) allowed
%Control: page (1) range
%Control: year (1) truncated
%Control: production of eprint (0) enabled

\end{document}